\documentclass{aa}

\usepackage[colorlinks=true, allcolors=blue]{hyperref}
\usepackage{graphicx} % for \includegraphics
\usepackage[subrefformat=parens]{subcaption} % for subfigures
\usepackage[varg]{txfonts} % for A&A fonts
\usepackage{mathrsfs} % for \mathscr
\usepackage{cleveref} % for \cref
\usepackage{csquotes} % for \enquote

\makeatletter % interpret @ in commands
\renewcommand*{\aa@journalname}{} % see aa.cls
\renewcommand*{\aa@manuscriptname}{} % see aa.cls
\makeatother

\newcommand\kmin{{k_\text{min}}}
\newcommand\kmax{{k_\text{max}}}
\newcommand\ellmin{{\ell_\text{min}}}
\newcommand\ellmax{{\ell_\text{max}}}
\newcommand\EB{Einstein–Boltzmann}

\newcommand\scrH{\mathscr{H}}
\newcommand\diff[1]{\mathrm{d}{#1}}

\begin{document}

\title{Chebyshev interpolation in \EB{} codes}
\authorrunning{Sletmoen, H.}
\author{Herman Sletmoen\thanks{Email: \texttt{herman.sletmoen@astro.uio.no}}}
\institute{Institute of Theoretical Astrophysics, University of Oslo, PO Box 1029 Blindern, 0315 Oslo, Norway}
\date{Received XX / Accepted XX}
\abstract{
	\EB{} codes compute theoretical predictions of cosmological models
	and rely heavily on interpolation in their independent variables:
	time $\tau$, wavenumber $k$ and multipole $\ell$.
	We give a practical summary of interpolation with Chebyshev polynomials,
	which converges rapidly for smooth functions and thus pairs naturally with approximation-free \EB{} codes.
	By solving the perturbations and line-of-sight integrals at Chebyshev nodes in $k$ and $\ell$,
	we show that Chebyshev polynomials interpolate to higher precision than traditional cubic splines from fewer explicit solutions.
	On a set of example spectra for matter and the cosmic microwave background (CMB),
	we find up to four orders of magnitude lower interpolation error using the same number of points.
	For a typical CMB temperature spectrum computed with interpolation in both $k$ and $\ell$,
	Chebyshev polynomials converge to $10^{-4}$--$10^{-5}$ relative error with only 50--80 points per variable,
	while cubic splines approach $10^{-4}$ error with 200 points, translating to a $2.5\times$--$4\times$ speedup.
	The exact improvement depends on the target function
	and is generally more dramatic at high precision levels.
	Standard Chebyshev $\ell$-interpolation needs line-of-sight integrals generalized to non-integer $\ell$,
	but we show a way to avoid this by rounding the nodes to integers.
	Chebyshev interpolation is implemented in SymBoltz,
	which is available at \url{https://github.com/hersle/SymBoltz.jl}.
}

\keywords{methods: numerical – cosmology: theory}

\maketitle

\nolinenumbers

\section{Introduction}
\label{sec:intro}

\EB{} codes, such as SymBoltz \citep{sletmoenSymBoltzjlSymbolicnumericApproximationfree2026a}, CLASS \citep{blasCosmicLinearAnisotropy2011b} and CAMB \citep{lewisEfficientComputationCMB2000b},
are fundamental tools for computing theoretical predictions from cosmological models.
They solve large and stiff ordinary differential equations (ODEs) and many line-of-sight integrals of the background and perturbation equations.
This outputs source functions $S(\tau)$ and $S(\tau, k)$ and transfer functions $\Delta_\ell(k)$
of conformal time $\tau$, perturbation wavenumber $k$ and angular multipole $\ell$.

Interpolation is one of the most effective optimizations in these codes.
Instead of evolving every $k$ and $\ell$ explicitly,
they do this only on coarse grids and later interpolate to more values.
This works because the output is smooth and saves time, but introduces interpolation errors.
Codes have default settings tuned to balance performance and precision,
but they slow down significantly with finer precision parameters that sample more points.
In the literature,
interpolation in parameter space with polynomials or neural network \enquote{emulators} has received much attention
\citep[e.g.][]{fendtPicoParametersImpatient2007,albersCosmicNetPhysicsdrivenImplementation2019a,manciniCOSMOPOWEREmulatingCosmological2022,boniciCapsejlEfficientAutodifferentiable2024a},
while comparatively little attention has been paid to interpolation in the independent variables.

Since the pioneering COSMICS code \citep{maCosmologicalPerturbationTheory1995},
interpolation has been done with cubic splines.
While solid for general-purpose interpolation,
they are outperformed by interpolation with Chebyshev polynomials \citep{trefethenApproximationTheoryApproximation2019}.
Cubic splines interpolate a function through $N$ points with $n=N-1$ piecewise cubic polynomials and an error that falls algebraically as $n^{-4}$.
In contrast, Chebyshev interpolation fits a single $n$-th order polynomial through all points with an error that may drop exponentially as $\rho^{-n}$ for some $\rho$.
This interpolates to higher precision from fewer samples.
It is also simple to implement with the barycentric formula \citep{berrutBarycentricLagrangeInterpolation2004},
and unlike equispaced polynomial interpolation it is stable \citep{trefethenSixMythsPolynomial}.
However, the remarkably fast convergence depends on the function being smooth,
as is often true in physics.

Traditional \EB{} codes such as CAMB and CLASS use approximation schemes to switch between different equations at different $\tau$, $k$ and $\ell$,
particularly in the perturbations \citep[notably Figure 10]{blasCosmicLinearAnisotropy2011b}.
This is faster, but complicates both the physics and numerics.
Moreover, it breaks smoothness by introducing kinks at the switches.
Cubic splines are good for interpolating such functions because kinks affect their piecewise polynomials only locally.
Chebyshev interpolation instead uses global polynomials, and a non-smooth feature at one point can slow convergence over the entire domain.

Recently, approximation-free \EB{} codes such as SymBoltz \citep{sletmoenSymBoltzjlSymbolicnumericApproximationfree2026a}, DISCO-EB \citep{hahnDISCODJDifferentiableEinsteinBoltzmann2024} and PyCosmo \citep{refregierPyCosmoIntegratedCosmological2018} have emerged.
They integrate the same equations at all times to obtain fully smooth solutions, but spend more effort per point.
In this paper, we show that approximation-free codes can leverage this smoothness with the rapid convergence of Chebyshev interpolation for performance and precision.
We implement this feature in SymBoltz\footnote{\url{https://hersle.github.io/SymBoltz.jl}}.

\section{Computations in \EB{} codes}
\label{sec:sources}

The equations in \EB{} codes are solved in stages for $\tau$, $k$ and $\ell$,
each interpolating solutions of earlier stages.
Background functions $S(\tau)$ are interpolated in $\tau$;
perturbation sources $S(\tau, k)$ in $\tau$ and $k$;
and angular spectra $C_\ell$ in $\ell$ \citep[e.g.][]{callinHowCalculateCMB2006c}.

First, background and thermodynamics equations are solved for variables $S(\tau)$ that depend only on $\tau$.
For example, the Friedmann, continuity and Peebles equations are solved for the Hubble function $\scrH(\tau)$, free electron fraction $X_e(\tau)$, optical depth $\kappa(\tau)$, visibility function $v(\tau)$ and lookback time $\chi(\tau) = \tau_0 - \tau$.

\begin{figure*}
	\centering
	\begin{subfigure}{0.24\textwidth}
		\includegraphics[width=\textwidth]{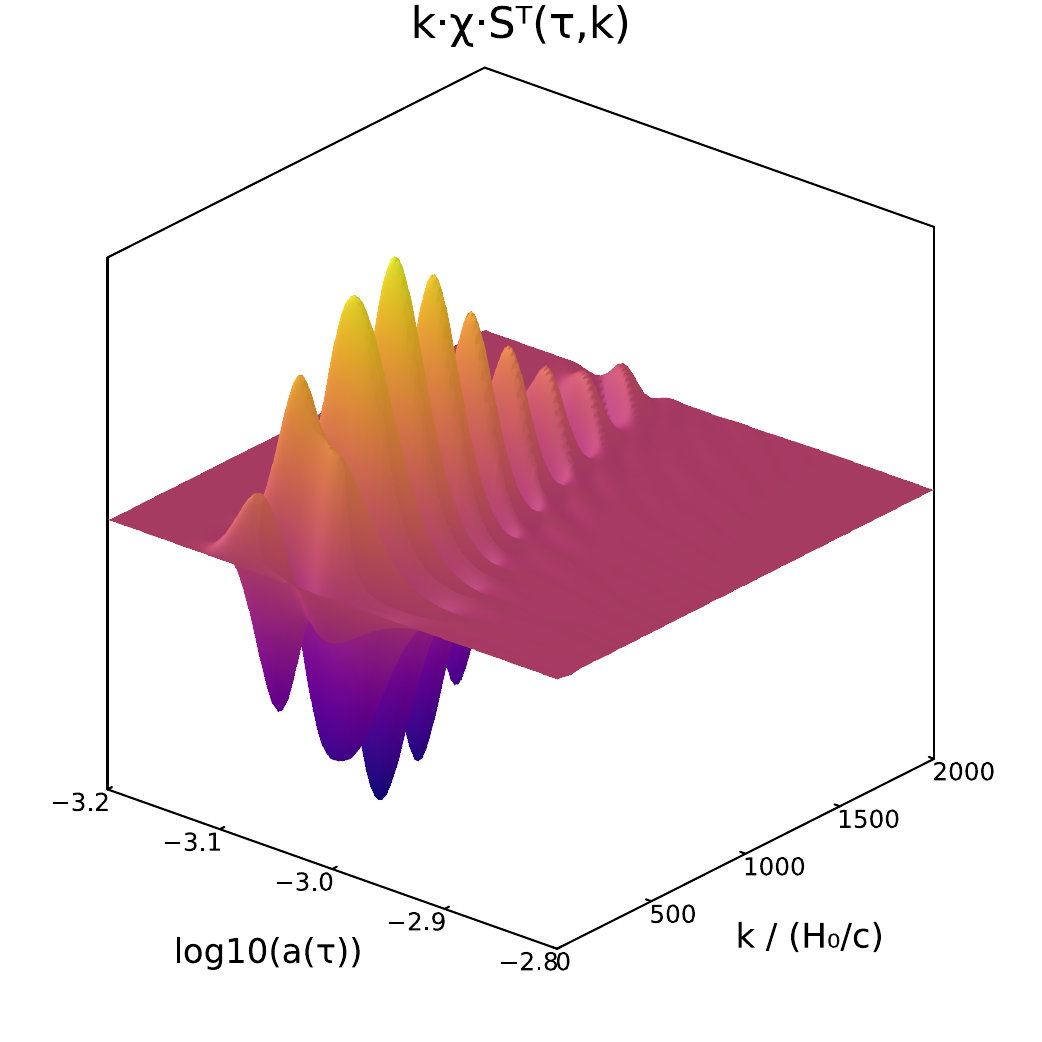}
		\subcaption{\label{fig:sourceT}Temperature}
	\end{subfigure}
	\hfill
	\begin{subfigure}{0.24\textwidth}
		\includegraphics[width=\textwidth]{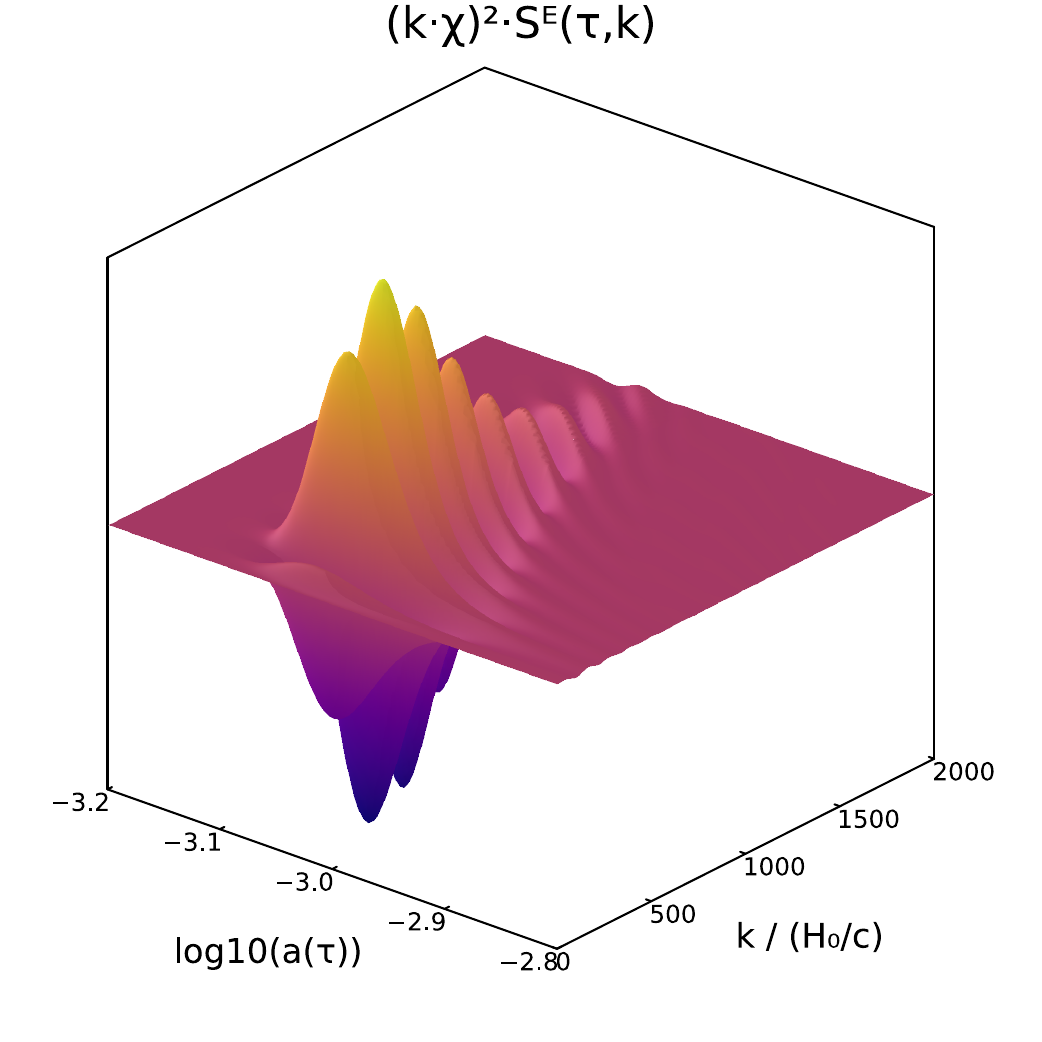}
		\subcaption{\label{fig:sourceE}Polarization}
	\end{subfigure}
	\hfill
	\begin{subfigure}{0.24\textwidth}
		\includegraphics[width=\textwidth]{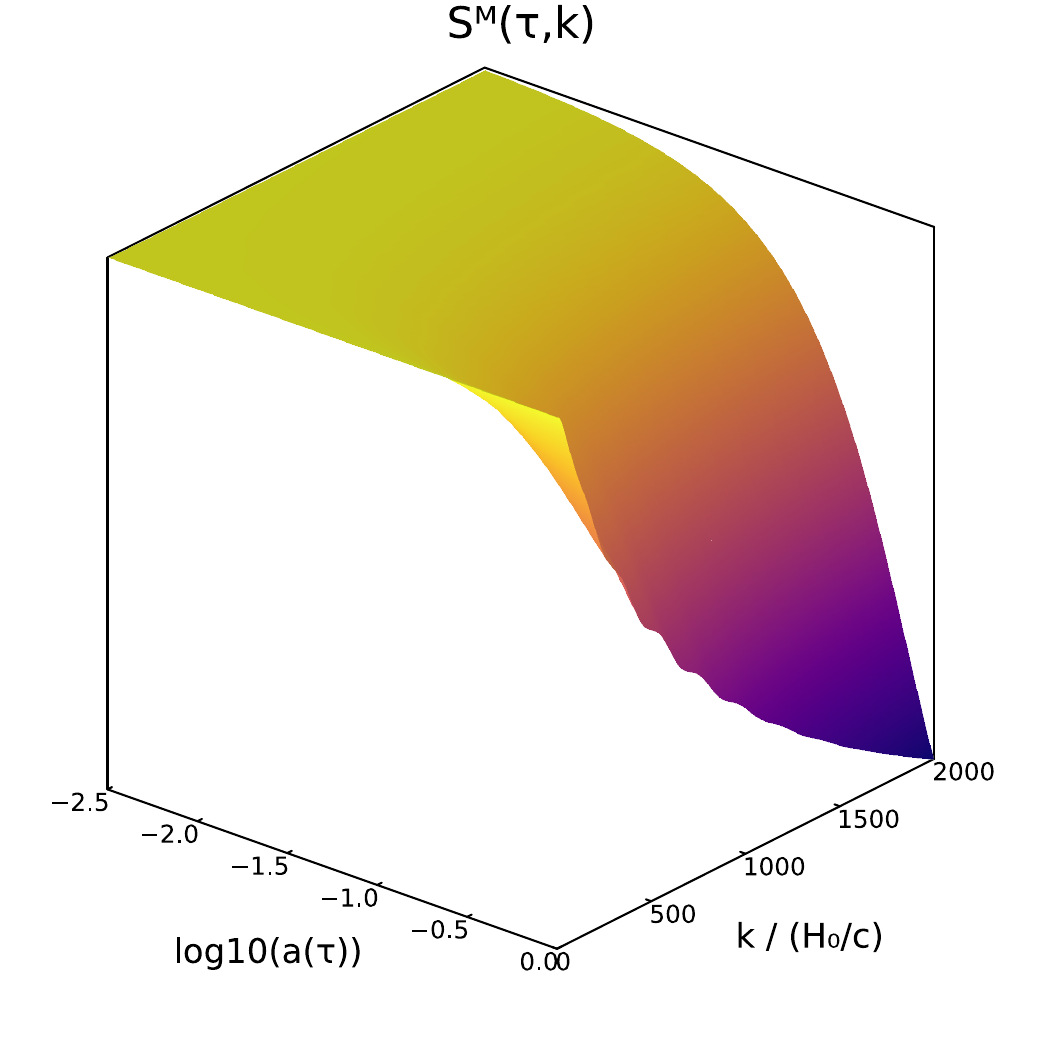}
		\subcaption{Matter}
	\end{subfigure}
	\hfill
	\begin{subfigure}{0.24\textwidth}
		\includegraphics[width=\textwidth]{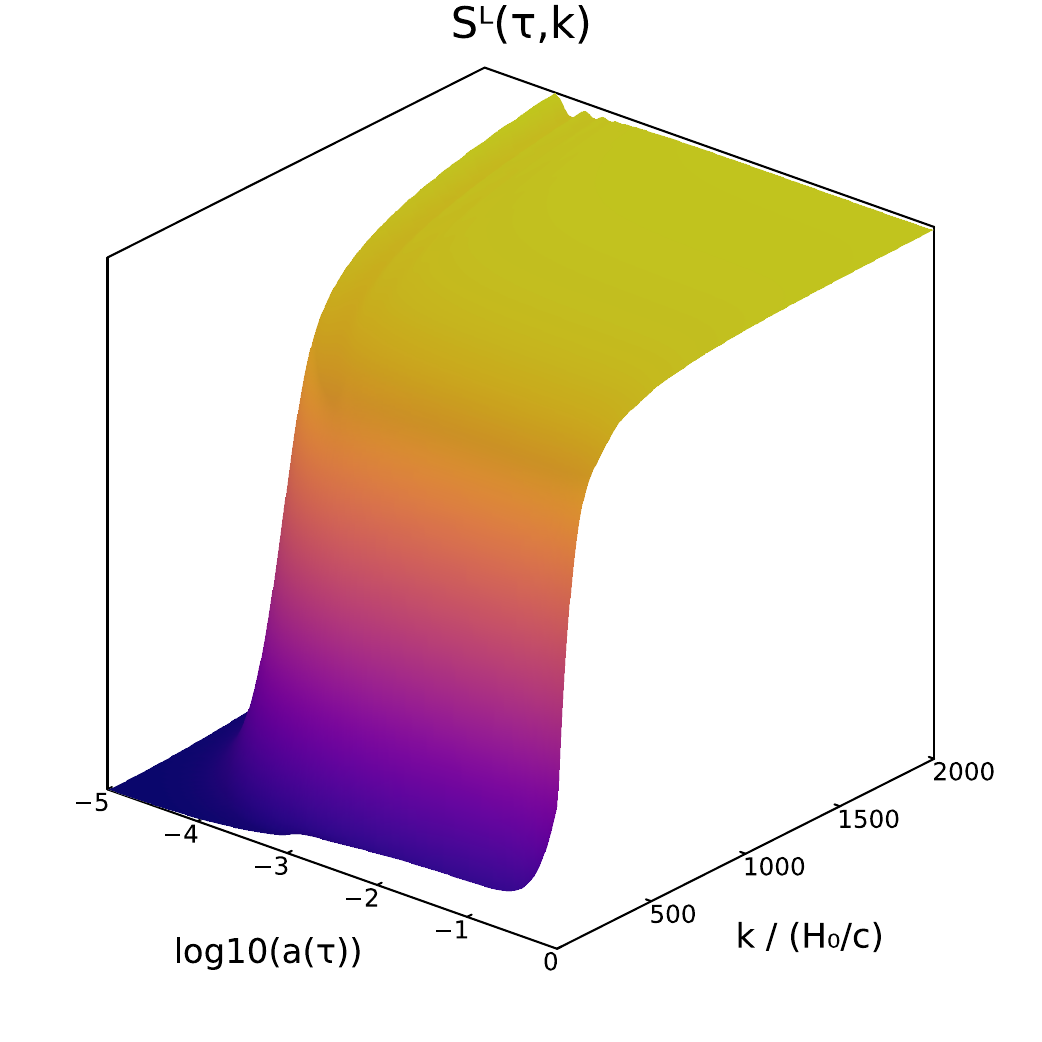}
		\subcaption{Lensing}
	\end{subfigure}
	\caption{%
		\label{fig:sources}
		Examples of the source functions \eqref{eq:sources} for temperature, polarization, matter and lensing (up to some normalization).
		The first two are shown around recombination,
		and the last two throughout cosmic time.
		The vertical scale is linear in arbitrary units.
	}
\end{figure*}

Second, large and stiff perturbation ODEs are integrated in $\tau$ for independent $k$-modes.
This must interpolate the background solution in $\tau$.
It produces source functions $S(\tau, k)$ for CMB temperature ($T$), polarization ($E$), lensing ($L$) and matter overdensity ($M$; see \cref{fig:sources} and \cite{sletmoenSymBoltzjlSymbolicnumericApproximationfree2026a} for full variable definitions):%
\begin{subequations}
	\begin{align}
		S^T    & = v \, \Bigg(\frac{\delta_\gamma}{4} + \Psi + \frac{\Pi_\gamma}{16}\Bigg) + e^{-\kappa} \, (\Psi + \Phi)^\prime + (v u_b)^\prime + \frac{3\big(v \Pi_\gamma\big)^{\prime\prime}}{16 k^2}, \\
		S^E    & = \frac{3 v \Pi_\gamma}{16 \, (k \chi)^2}, \\
		S^L    & = (\Psi+\Phi) \, \frac{\chi_\text{rec}}{(\chi-\chi_\text{rec}) \chi}, \\
		S^M    & = \delta_m + 3 (1+w_m) \, \frac{\scrH}{k^2} \, \theta_m .
	\end{align}
	\label{eq:sources}%
\end{subequations}

Third, line-of-sight integrals
\citep{seljakLineofSightIntegrationApproach1996}%
\begin{equation}
    \Delta^A_\ell(k) = \int_{\tau_i}^{\tau_0} \diff{\tau} \, S^A(\tau, k) j_\ell(k \chi)
    \label{eq:los}
\end{equation}
are evaluated for a given source $S^A(\tau, k)$,
which must be interpolated in $\tau$ and $k$.
In turn, this is used to compute angular spectra%
\begin{equation}
    C_\ell^{AB} \propto \int_\kmin^\kmax \diff{k} \, k^2 P_0(k) \Delta^A_\ell(k) \Delta^B_\ell(k) \quad \text{(up to an $\ell$-scaling)}.
    \label{eq:spectra}
\end{equation}

Most $S(\tau, k)$ are smooth in $k$.
But the spherical Bessel function $j_\ell(k\chi)$ makes $\Delta_\ell(k)$ oscillate rapidly in $k$,
so the integrand of $C_\ell$ must be sampled at many $k$.
A standard trick is to solve perturbations explicitly for only $10^2$--$10^3$ $k$-modes,
and then interpolate $S(\tau, k)$ to $10^3$--$10^4$ values before line-of-sight integration.
This is $10\times$--$100\times$ faster than evolving the perturbations for all $k$.

Most $C_\ell$ are also smooth in $\ell$.
Similarly, it is faster to interpolate $C_\ell$ from some $\ell$ to every integer.
This holds particularly in curved universes, which need hyperspherical Bessel functions.

\section{Chebyshev polynomial interpolation}
\label{sec:chebyshev}

Any function $f(x)$ is interpolated through $N$ points $(x_i, f_i)$ by a polynomial $p_n(x)$ of order $n=N-1$:%
\begin{equation}
	f(x) \approx p_n(x) = \sum_i f_i \, L_i(x)
	\quad \text{with} \quad
	L_i(x) = \frac{\prod_{j \neq i} \, (x-x_j)}{\prod_{j \neq i} \, (x_i-x_j)} .
\label{eq:interppoly}
\end{equation}
The Lagrange polynomials satisfy $L_i(x_j) = \delta_{ij}$, so $p_n(x_i) = f_i$.
We work with the standard domain $x \in [-1, 1]$,
and map interpolation over arbitrary $y \in [a, b]$ with $x(y) = ( 2y-b-a ) / (b-a)$.

The polynomial interpolation error is bounded by
\begin{equation}
	\big\lvert f(x) - p_n(x) \big\rvert \leq \frac{M_{n+1}}{(n+1)!} \, \lvert \omega(x) \rvert,
	\quad
	\omega(x) = \prod_i (x-x_i),
\label{eq:error}
\end{equation}
where $M_{n} = \max \lvert f^{(n)}(x) \rvert$ on $x\in[-1,1]$.
Thus, the points should be placed in a way that minimizes the nodal polynomial $\lvert \omega(x) \rvert$.

Unintuitively, it is a bad idea to use uniformly spaced $x_i$.
Then $\omega(x)$ decays very slowly with $n$ near the edges and is easily outrun by $M_{n+1}/(n+1)!$ as $n \rightarrow \infty$.
This is called Runge's phenomenon and causes violent oscillations demonstrated in \cref{fig:recequi}.

Two particularly excellent ways to place the points are at the Chebyshev nodes of the first or second kind
\citep{trefethenApproximationTheoryApproximation2019}:
\begin{subequations}
\begin{align}
	x_i^{(1)} &= \cos \left( \frac{2i+1}{2n+2} \pi \right) && (i = 0,\, 1,\, \ldots\, n), \label{eq:chebnodes1} \\
	x_i^{(2)} &= \cos \left( \frac{i}{n} \pi \right) && (i = 0,\, 1,\, \ldots\, n). \label{eq:chebnodes2}
\end{align}
\label{eq:chebnodes}%
\end{subequations}
The points $x_i^{(1)}$ give the best possible bound $\lvert \omega(x) \rvert \leq 2^{-n}$ for all $x$.
Meanwhile, $x_i^{(2)}$ bound $\lvert \omega(x) \rvert \leq 2^{-n+1}$ but include the endpoints $x = \pm 1$,
which is sometimes more or less practical.
In practice,
$x_i^{(1)}$ and $x_i^{(2)}$ and any other set of points distributed with a similar $(1-x^2)^{-1/2}$-proportional density are effective.
Here we use $x_i^{(2)}$.

The most attractive property of Chebyshev interpolation is that it converges faster the smoother $f(x)$ is \citep{trefethenSixMythsPolynomial}:
\begin{subequations}
\begin{align}
	\big\lvert f(x) - p_n(x) \big\rvert &\leq C \, \rho^{-n} \quad \text{if $f(x)$ is analytic,} \label{eq:geometric} \\
	\big\lvert f(x) - p_n(x) \big\rvert &\leq C \, n^{-\nu} \quad \text{if $f(x)$ is differentiable $\nu$ times.} \label{eq:algebraic}
\end{align}
\label{eq:convergence}%
\end{subequations}
This is called \enquote{geometric} and \enquote{algebraic} convergence, respectively.
Here $C$ and $\rho > 1$ are constants that depend on $f(x)$.

Evaluating $p_n(x)$ is simple, fast and robust through the barycentric interpolation formula \citep{berrutBarycentricLagrangeInterpolation2004}:%
\begin{equation}
	p_n(x) = \sum_i \frac{w_i}{x-x_i} f_i \,\,\,\Bigg/\,\,\, \sum_i \frac{w_i}{x-x_i}.
\label{eq:bary}
\end{equation}
This involves the point-dependent interpolation weights
\begin{equation}
	w_i = \frac{1}{\omega^\prime(x_i)} = \frac{1}{\prod_{j \neq i} (x_i - x_j)}.
\label{eq:weights}
\end{equation}
They can be precomputed for a general scheme,
but for the Chebyshev nodes \eqref{eq:chebnodes} one can instead just use the exact weights%
\begin{subequations}
\begin{align}
	w_i^{(1)} &= (-1)^i \sin \left( \frac{(2i+1)}{2n+2} \pi \right), \label{eq:chebweights1} \\
	w_i^{(2)} &= \begin{cases} (-1)^i/2 & \text{if } i = 0 \text{ or } i = n, \\ (-1)^i & \text{otherwise}. \end{cases} \label{eq:chebweights2}
\end{align}
\label{eq:chebweights}%
\end{subequations}
Barycentric interpolation is robust and uses only $O(n)$ operations per $x$.
Despite suspicious divisions by $x-x_i$, it is stable in floating point arithmetic even as $x \rightarrow x_i$,
and needs only a simple exception to return $f_i$ if $x = x_i$ exactly \citep{highamNumericalStabilityBarycentric2004}.
In contrast, the Lagrange form \eqref{eq:interppoly} is unstable and uses $O(n^2)$ operations per $x$.

Our implementation creates a \texttt{ChebyshevInterpolator} from an interval $[a, b]$ and polynomial degree $n$,
computes and transforms the nodes \eqref{eq:chebnodes} to $[a, b]$,
and then evaluates the barycentric formula \eqref{eq:bary} with weights \eqref{eq:chebweights} and any function values $f(x_i)$.

An alternative implementation expands $p_n(x) = \sum_m c_m T_m(x)$ in a series of Chebyshev polynomials $T_m(x)$.
A Discrete Cosine Transform converts between the coefficients $c_m$ and function values $f_i$.
This lets us inspect the convergence of $\lvert c_m \rvert$ with $m$,
and thus convergence of the full polynomial $p_n(x)$.
It is not necessary, but is useful for error analysis and described in \cref{sec:coeffs}.

In contrast, piecewise interpolation methods join several low-order polynomials and converge more slowly.
Cubic splines are popular for smooth data and interpolate with an error that falls as $n^{-4}$.
They can struggle near the ends, where splines assume boundary conditions that need not match the data.
In return, splines have better stability on arbitrary $x$-grids and respond locally to changes in the data.
Splines are also faster to evaluate,
but this is irrelevant when the cost of evaluating $f(x)$ dominates.

Thus, Chebyshev interpolation is very attractive when:%
\begin{itemize}
\item $f(x)$ is smooth, to get the rapid convergence \eqref{eq:convergence};
\item $f(x)$ is queryable, to sample at the needed points \eqref{eq:chebnodes};
\item $f(x)$ is expensive, to benefit most from fewer samples.
\end{itemize}
These conditions are seldom met by observed data.
But they often hold in theoretical applications such as \EB{} codes,
which model smooth physics with demanding computations and can choose sampling points freely.
In the next sections, we systematically consider interpolation in each independent variable:
time $\tau$, wavenumber $k$ and multipole $\ell$.

\section{Interpolation in time}
\label{sec:time}

We do \emph{not} use Chebyshev interpolation in $\tau$.
Nevertheless, we find it instructive to discuss this aspect in order to understand the tradeoffs compared to Chebyshev interpolation in $k$ and $\ell$.

The background and perturbations for each $k$ are ODEs $u^\prime(\tau) = f(u, \tau)$ in $\tau$.
They have inherently local features,
as the universe undergoes eras dominated by radiation, matter and dark energy,
and shorter events such as recombination (see \cref{fig:sourceT}) and reionization.
The ODEs are solved by adaptive methods that automatically adjust their time step to capture the local features.
This conflicts with the rigid Chebyshev node placement \eqref{eq:chebnodes}.

Moreover, ODE solvers feature specialized dense output methods for interpolating over time steps.
They are usually custom-tailored to the solver and optimized to maximize accuracy while reusing internal calculations of $f(u, \tau)$ at internal stages.
For example, SymBoltz uses the implicit \texttt{Rodas5P} ODE solver by default,
which has a specialized fourth-order interpolant \citep{steinebachConstructionRosenbrockWanner2023a}.
Even without specialized dense output,
the time derivative $u^\prime(\tau) = f(u, \tau)$ can always be computed analytically from the ODE definition and solution $u(\tau)$ \enquote{for free}.
One can then fall back to cubic Hermite interpolation,
which takes both $u(\tau)$ and $u^\prime(\tau)$ into account for improved accuracy.
We do not use Chebyshev interpolation to compete with an interpolation method that is intrinsic to this algorithm.

To go beyond the scope of this paper and take Chebyshev polynomials in time seriously,
one could use spectral ODE methods.
Instead of evolving $u(\tau)$,
they expand $u(\tau) = \sum_m a_m T_m(x(\tau))$ in Chebyshev polynomials $T_m(x(\tau))$
and derive equations for $a_m$ from the ODE
\citep[e.g.][]{choustikovOptimizingEvolutionPerturbations2023,ranaSpectralChebyshevApproximation2025}.

\section{Interpolation in wavenumber}
\label{sec:wavenumbers}

In $k$, however, Chebyshev interpolation is a good candidate.
The perturbation ODEs are independent for different $k$,
contain no analytical $k$-derivatives for Hermite interpolation,
and have a global nature in $k$ as all scales inherit the background at a time $\tau$.

Chebyshev $k$-interpolation consists of evolving the perturbation ODEs for the Chebyshev $k$-nodes \eqref{eq:chebnodes} of a wavenumber interval $[\kmin, \kmax]$.
Afterwards, any source $S(\tau, k)$ is computed at any $k$ by doing barycentric interpolation in $k$ over fixed $\tau$-slices.

\Cref{fig:sources} shows the four test source functions \eqref{eq:sources} that we interpolate.
We interpolate $k \chi \, S^T(\tau, k)$ and $(k\chi)^2 \, S^E(\tau, k)$,
as they are smoother and more similar in magnitude over $k$ than $S(\tau, k)$.

\begin{figure*}
	\centering
	\begin{subfigure}{0.32\textwidth}
		\includegraphics[width=\textwidth]{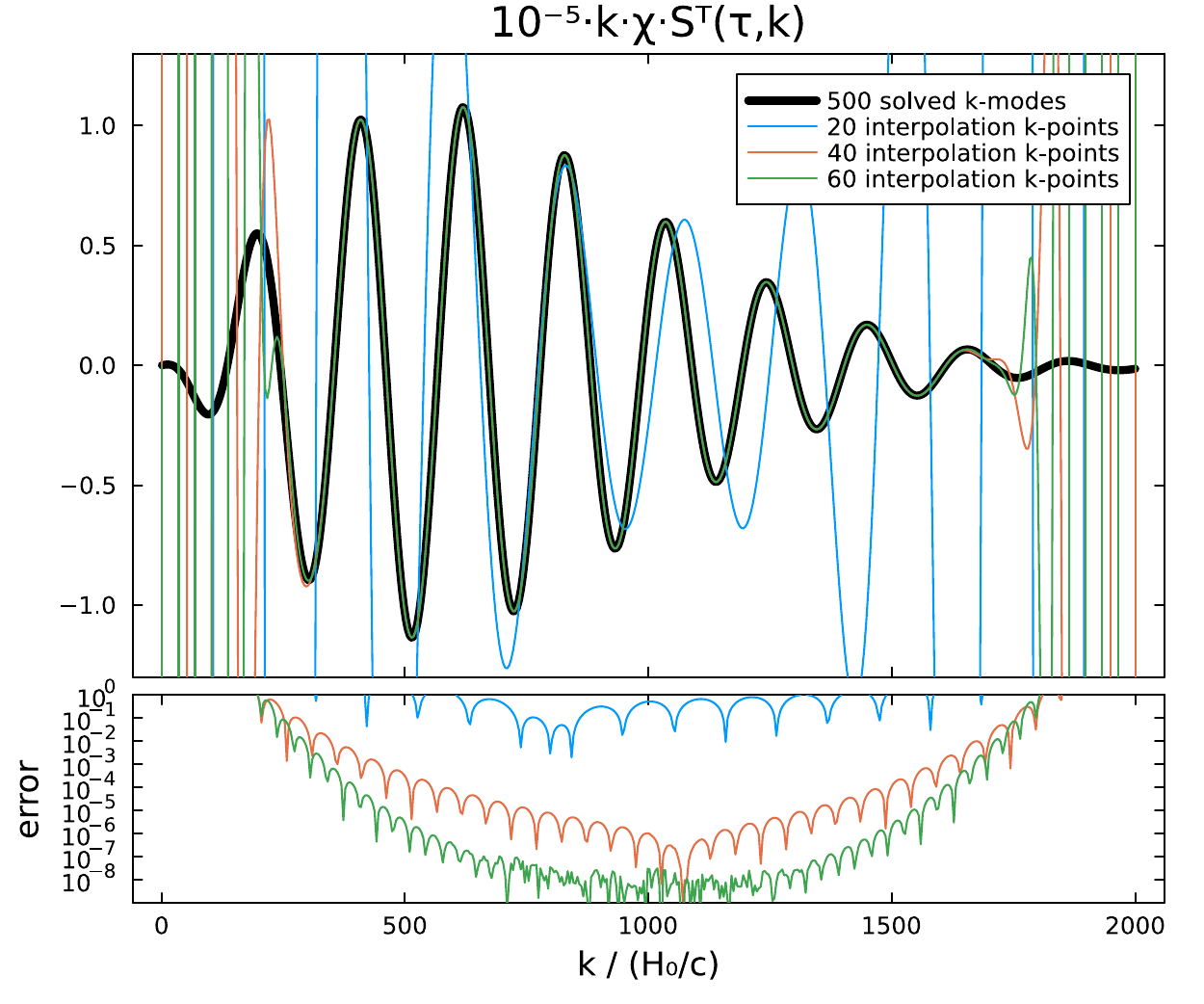}
		\subcaption{Equispaced polynomial interpolation}
		\label{fig:recequi}
	\end{subfigure}
	\hfill
	\begin{subfigure}{0.32\textwidth}
		\includegraphics[width=\textwidth]{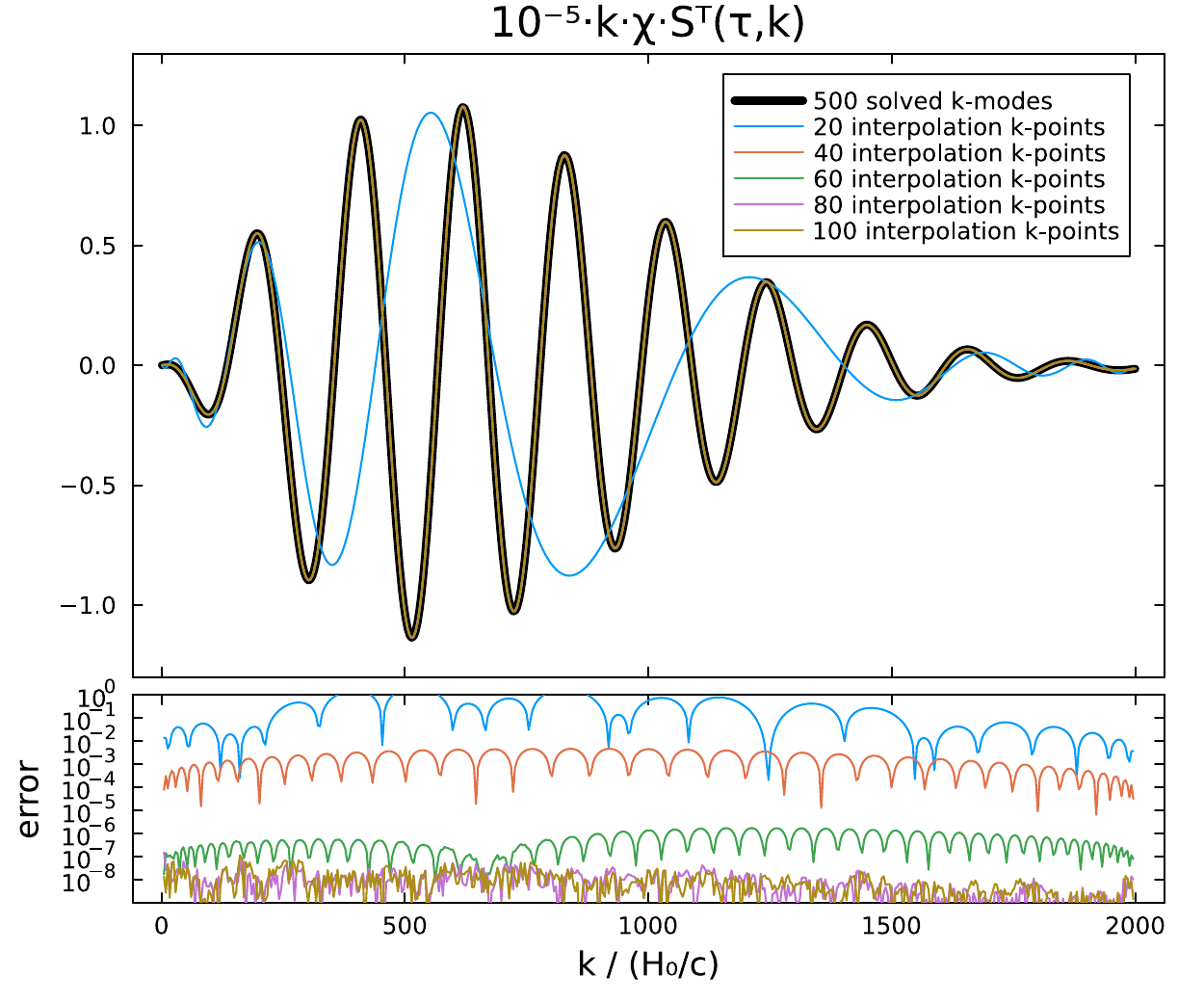}
		\subcaption{Chebyshev polynomial interpolation}
		\label{fig:reccheb}
	\end{subfigure}
	\hfill
	\begin{subfigure}{0.32\textwidth}
		\includegraphics[width=\textwidth]{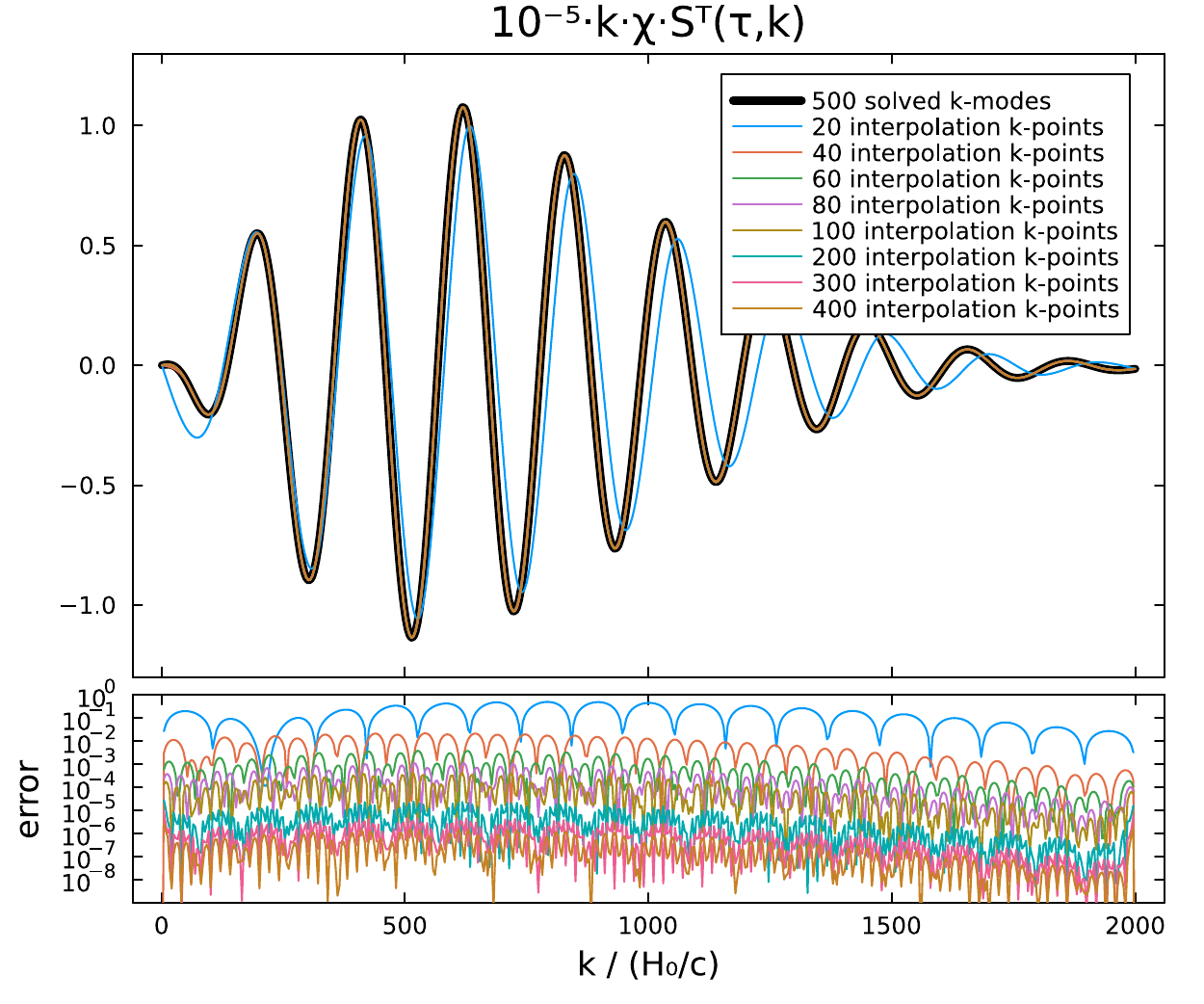}
		\subcaption{Equispaced cubic spline interpolation}
		\label{fig:reccubic}
	\end{subfigure}
	\caption{%
		\label{fig:rec}
		Temperature source in \cref{fig:sourceT} at recombination interpolated with \subref{fig:recequi} equispaced polynomials, \subref{fig:reccheb} Chebyshev polynomials and \subref{fig:reccubic} equispaced cubic splines.
		The interpolated values are compared to explicit solutions for 500 perturbation $k$-modes.
	}
\end{figure*}

\Cref{fig:rec} shows the temperature source interpolated in $k$ at recombination.
It demonstrates that Chebyshev interpolation and equispaced cubic splines are stable,
while Runge's phenomenon destabilizes equispaced polynomial interpolation.
At $N \gtrsim 40$, the Chebyshev interpolation error falls rapidly below the cubic spline error.
At $N = 80$, Chebyshev interpolation has fully converged to the numerical noise floor and is indistinguishable from solving all perturbation $k$-modes explicitly,
while cubic splines have 4 orders of magnitude higher error.
At $N = 400$, cubic splines have still not fully converged.
This is the fast convergence \eqref{eq:convergence} of Chebyshev interpolation for smooth functions.

\begin{figure*}
	\centering
	\includegraphics[width=\textwidth]{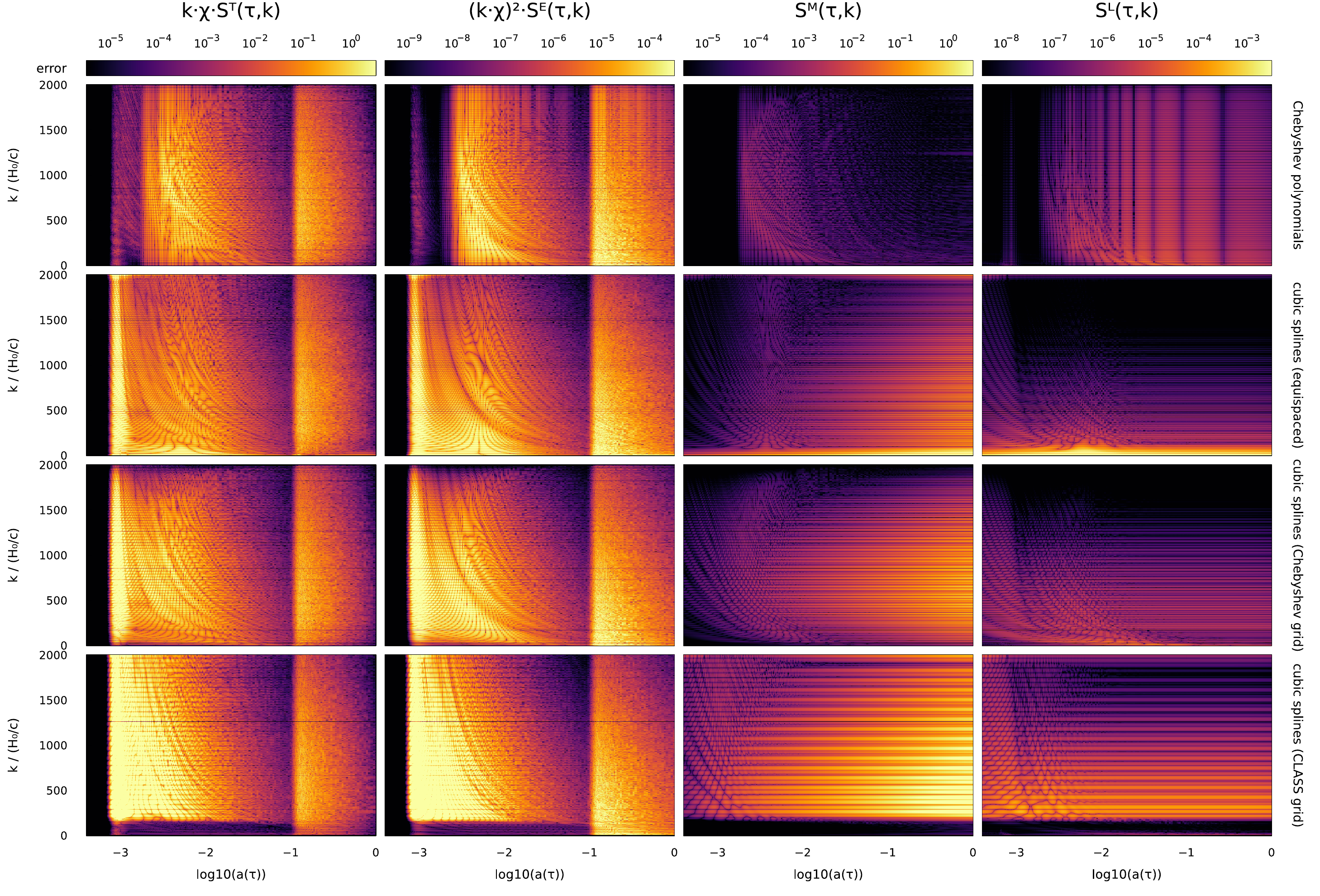}
	\caption{%
		\label{fig:heaterrs}
		Interpolation error of temperature, polarization, matter and lensing source functions in \cref{fig:sources} (left to right)
		interpolated in $k$ from 100 points using Chebyshev polynomials, equispaced cubic splines, cubic splines on a Chebyshev grid and cubic splines on a CLASS-like grid (top to bottom).
		As in \cref{fig:rec}, the error is computed relative to 500 explicitly solved perturbation $k$-modes.
	}
\end{figure*}

\Cref{fig:heaterrs} generalizes the comparison to all four sources \eqref{eq:sources} and the full $(\tau, k)$-space for $N = 100$.
For the temperature and polarization sources,
Chebyshev interpolation is 4 orders of magnitude more accurate than uniform cubic splines in a band around recombination at $a(\tau) \approx 10^{-3}$,
where they are most important to capture.
For the matter and lensing sources,
Chebyshev interpolation is also more accurate and distributes the error uniformly over $k$.
Note that cubic splines struggle particularly at the boundaries of the interpolation interval.
This is because their natural boundary conditions assume that second derivatives are zero there,
which disagrees with the underlying function.

\Cref{fig:heaterrs} also shows cubic splines interpolated on the Chebyshev grid \eqref{eq:chebnodes}.
This only gives a lower error near the boundaries,
confirming that the advantage of Chebyshev interpolation over uniform cubic splines is not merely due to its sample locations.
In addition, the fourth row shows cubic splines with a non-uniform $k$-grid using CLASS' default strategy,
where $k$-modes are sampled with $25\times$ higher density for scales above the sound horizon at recombination than below it.%
\footnote{Using \texttt{k\_step\_sub=0.05} and \texttt{k\_step\_super=0.002} in CLASS v3.3.4, but downsampling the grid from around 500 to 100 $k$-points.}
This much denser sampling gives cubic splines the lowest error for $k \lesssim 200\,H_0/c$.
Meanwhile, Chebyshev interpolation gives low average error throughout the domain with straightforward sampling.

\begin{figure*}
	\centering
	\includegraphics[width=\textwidth]{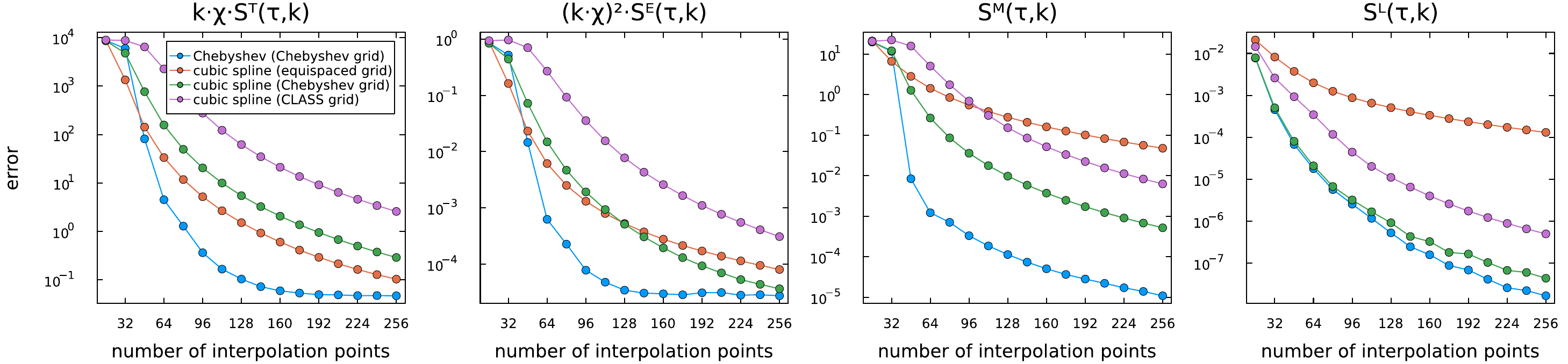}
	\caption{
		\label{fig:convergence}
		Root-mean-square of the interpolation error over $(\tau, k)$
		using $k$-interpolation with Chebyshev polynomials, uniform cubic splines, cubic splines with Chebyshev points and cubic splines with CLASS-like $k$-sampling (as in \cref{fig:heaterrs})
		as the number of samples increases.
	}
\end{figure*}

\Cref{fig:convergence} compresses the error over $(\tau, k)$ to one root-mean-square number,
and shows its convergence with more samples.
The ranking of the methods depends on the source, and also crosses over with $N$.
By this metric, Chebyshev interpolation achieves the lowest error for every source and $N \gtrsim 48$.
This further supports that it is a better overall interpolation method when measured over the entire $k$-domain.

\begin{figure*}
	\centering
	\includegraphics[width=\textwidth]{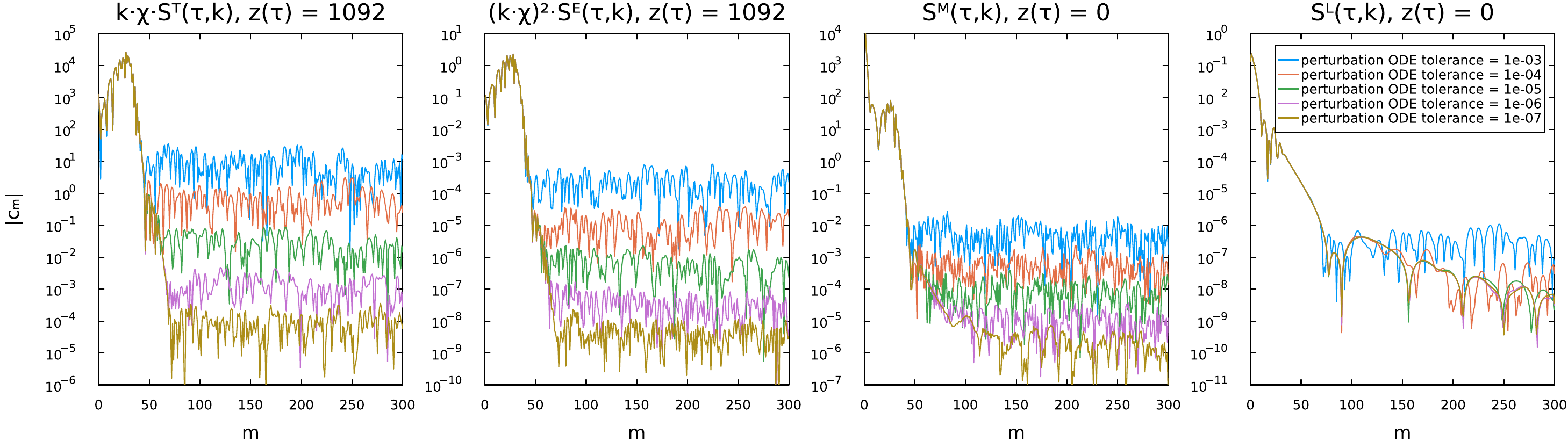}
	\caption{%
		\label{fig:coeffs}
		Coefficient magnitudes $\lvert c_m \rvert$ in the Chebyshev polynomial expansion $S(\tau, k) \approx \sum_{m=0}^{300} c_m(\tau) T_m(x(k))$ of the source functions in \cref{fig:sources}.
		The temperature and polarization sources are evaluated at recombination,
		and the matter and lensing sources today.
		The noise floor for the interpolation error is lowered by stricter ODE solver tolerances for the perturbations.
	}
\end{figure*}

\Cref{fig:coeffs} studies the decaying coefficients $c_m$ in the Chebyshev series expansion $p_n(x) = \sum_m c_m T_m(x)$ of the interpolant.
As \cref{sec:coeffs} explains, this is an equivalent Chebyshev interpolation method that offers a useful diagnostic:
convergence of $\lvert c_m \rvert$ indicates convergence of $p_n(x)$.
It has no direct counterpart with splines.
The convergence undergoes three phases (here for $S^T$):%

\begin{itemize}
\item
For $0 \lesssim m \lesssim 25$,
the interpolant is still resolving the features of the source function and $\lvert c_m \rvert$ has not yet begun to converge.

\item
For $25 \lesssim m \lesssim 75$,
the interpolant has captured the main features and $\lvert c_m \rvert$ decays rapidly.
The steepest fall is a line in the log-linear plot,
indicating geometric convergence \eqref{eq:geometric}.

\item
For $m \gtrsim 75$,
the interpolant and $\lvert c_m \rvert$ fully converge to the numerical noise floor.
Smaller ODE solver tolerances lower the precision floor.
This matches the convergence with 60--80 points in \cref{fig:reccheb},
and there is little gain in using more points.
\end{itemize}
In principle, the code could use this decay to choose $n$ adaptively.

Thus, we find Chebyshev interpolation in $k$ to be a better overall method for the whole $k$-domain.
At high precision levels, it interpolates with orders of magnitude lower error than cubic splines from fewer explicit solutions to the perturbation ODEs.

\section{Interpolation in multipole}
\label{sec:multipoles}

Angular power spectra $C_\ell$ are also smooth functions of the multipole $\ell$.
Instead of computing $C_\ell$ explicitly with line-of-sight integration for every $\ell$,
\EB{} codes do this only for some $\ell$ and then interpolate to all desired integer $\ell$.
This again makes Chebyshev interpolation a good candidate.

However, the discrete nature of $\ell$ raises a difficulty.
As $C_\ell$ is physically interpreted only at integer $\ell$,
other codes interpolate from an integer $\ell$-grid (e.g. using every low $\ell$ and every $30$th high $\ell$).
This works with cubic splines because their stability is better for arbitrary grid configurations.
But it is incompatible with standard Chebyshev interpolation,
because the node placement \eqref{eq:chebnodes} forces the interpolation grid to both integer and non-integer $\ell$.
This was not a concern in \cref{sec:wavenumbers},
where sources vary continuously with $k$.
We show two different ways to solve this problem.

The first way is to proceed with standard Chebyshev interpolation,
but generalize the line-of-sight integrals \eqref{eq:los} and spherical Bessel functions $j_\ell(x)$ to arbitrary (integer and non-integer) $\ell$.
This is unproblematic from a physical point of view, as $j_\ell(x)$ is the solution that is regular at $x = 0$ of the ODE
\begin{equation}
	x^2 \frac{\mathrm{d}^2 j_\ell}{\mathrm{d} x^2} + 2x \frac{\mathrm{d} j_\ell}{\mathrm{d} x} + \left( x^2 - \ell(\ell+1) \right) j_\ell = 0.
\label{eq:besselode}
\end{equation}
This is continuous and smooth in both $\ell$ and $x$, so interpolation from non-integer to integer $\ell$ converges.
The important part is not the precise value of $j_\ell(x)$ for non-integer $\ell$,
but that there is a continuation of $j_\ell(x)$ between integer $\ell$ that is smooth.%
\footnote{Extending $j_\ell(x)$ to continuous $\ell$ is analogous to analytically continuing the factorial $n!$ to the gamma function $\Gamma(n+1)$, for example.}
Cubic spline interpolation between integer-only $\ell$ also relies on this property.
When a standard \texttt{ChebyshevInterpolator} is used to interpolate over $\ell \in [\ellmin, \ellmax]$,
SymBoltz follows this path by evaluating $j_\ell(x)$ at arbitrary $\ell$ with the Bessels.jl\footnote{\url{https://github.com/JuliaMath/Bessels.jl}} package.

The second way is to compute $j_\ell(x)$ only at integer $\ell$,
but round Chebyshev nodes to their nearest integers.
These non-standard points break the Chebyshev weights \eqref{eq:chebweights} and coefficients \eqref{eq:coeffsdiscrete},
but barycentric interpolation is still straightforward using the general weights \eqref{eq:weights}.
This near-Chebyshev method is robust,
as polynomial interpolation is stable for any points with density similar to the Chebyshev nodes \eqref{eq:chebnodes}.
It also helps that $C_\ell$ is needed for 1000s of multipoles,
which mitigates discreteness effects when rounding 10s--100s of interpolation points.
SymBoltz implements a \texttt{ChebyshevIntegerInterpolator} with this strategy,
and errors if the requested order is so high that multiple nodes round to the same $\ell$.

\begin{figure*}
	\centering
	\begin{subfigure}{0.32\textwidth}
		\includegraphics[width=\textwidth]{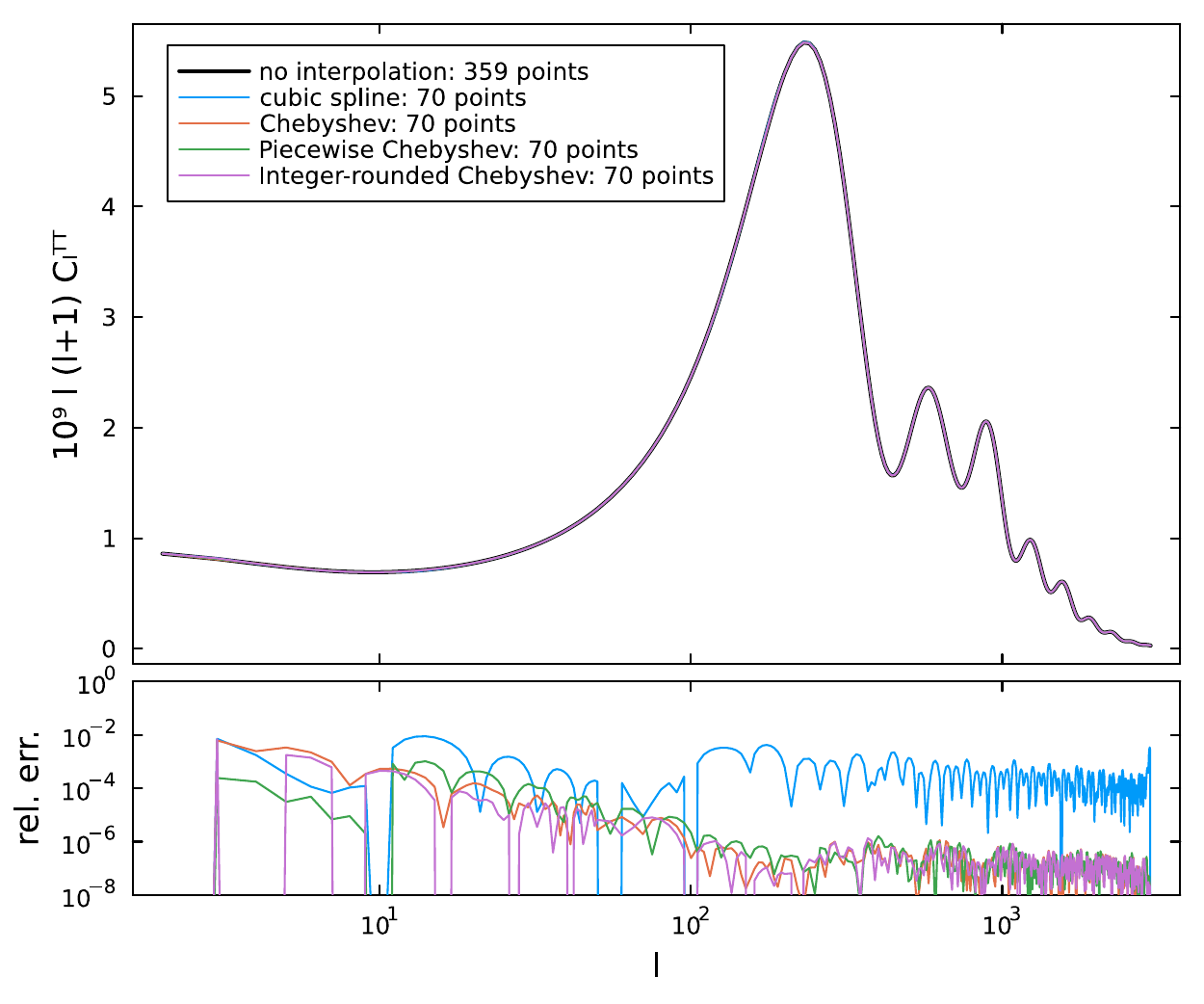}
		\subcaption{Temperature ($C_\ell^\mathrm{TT}$)}
		\label{fig:cmbTT}
	\end{subfigure}
	\hfill
	\begin{subfigure}{0.32\textwidth}
		\includegraphics[width=\textwidth]{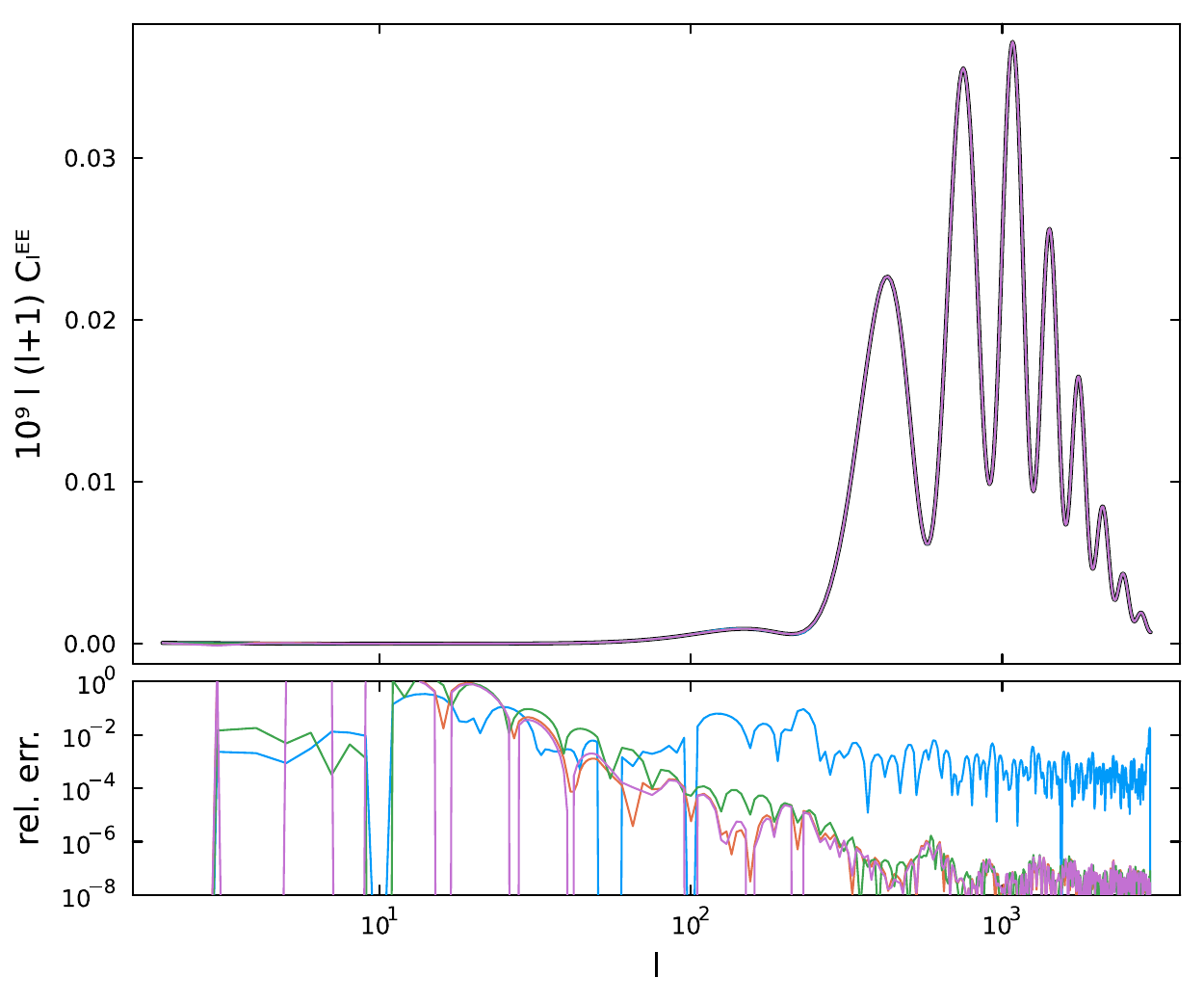}
		\subcaption{Polarization ($C_\ell^\mathrm{EE}$)}
		\label{fig:cmbEE}
	\end{subfigure}
	\hfill
	\begin{subfigure}{0.32\textwidth}
		\includegraphics[width=\textwidth]{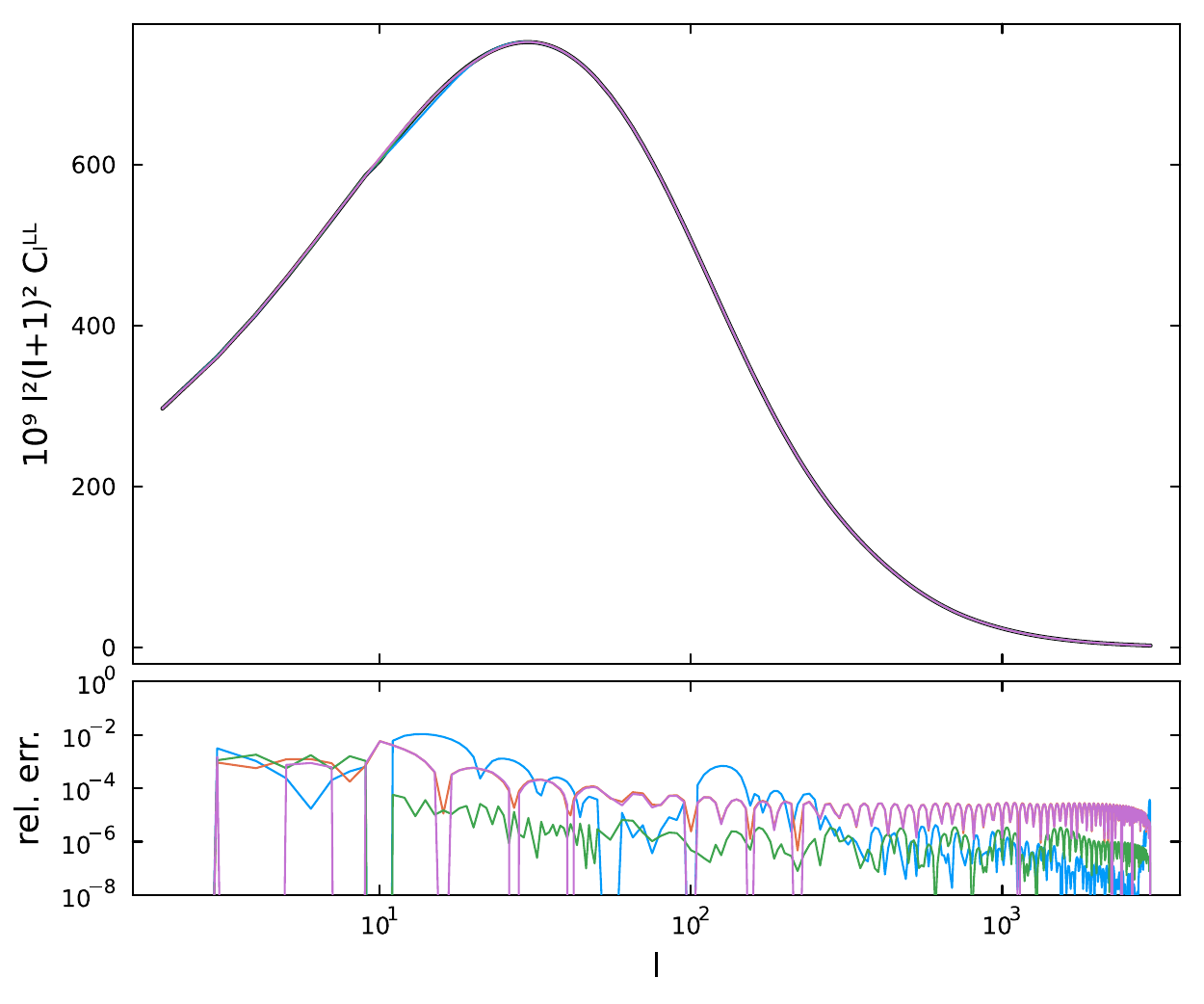}
		\subcaption{Lensing ($C_\ell^\mathrm{LL}$)}
		\label{fig:cmbLL}
	\end{subfigure}
	\caption{%
		\label{fig:cmb}
		CMB temperature, polarization and lensing spectra interpolated in $\ell$ with cubic spline versus Chebyshev polynomials.
		We interpolate $\ell^5 C_\ell$ as a function of $\ell$, which made all methods perform best.
		Cubic splines use 8, 9 and 55 samples on $\ell \in [2, 10]$, $[10, 100]$ and $[100, 3000]$, respectively.
		Standard Chebyshev interpolation fits a single 69th order polynomial through the Chebyshev nodes on $\ell \in [2, 3000]$.
		The piecewise variant joins a 5th order polynomial on $\ell \in [2, 10]$ with a 64th order polynomial on $\ell \in [10, 3000]$.
		The integer-rounded variant instead runs the degree-69 polynomial through the integers that are closest to the standard Chebyshev nodes.
		Every interpolated spectrum is compared to explicit line-of-sight integration for 359 $\ell$-values.
	}
\end{figure*}

\Cref{fig:cmb} compares $\ell$-interpolation performance of cubic splines versus Chebyshev polynomials.
To fix the computational cost, we use 70 $\ell$-points with each method.
For each method, we also require that the same $\ell$-grid interpolates all three spectra $C_\ell^\mathrm{TT}$, $C_\ell^\mathrm{EE}$ and $C_\ell^\mathrm{LL}$, so the precomputed $j_\ell(x)$ is common.

For $C_\ell^\mathrm{TT}$ and $C_\ell^\mathrm{EE}$, Chebyshev interpolation converges to 3--4 orders of magnitude smaller error for $\ell \gtrsim 100$ than cubic splines.
But for $C_\ell^\mathrm{LL}$, the cubic spline does better than a single Chebyshev polynomial.
This is because the Limber approximation switches on for $\ell \geq 10$ and makes a kink in $C_\ell$,
leading to only algebraic $n^{-1}$-convergence \eqref{eq:algebraic}.
The cubic spline then wins with $n^{-4}$ convergence because it responds locally,
so high $\ell$ is unaffected by the kink.
To fix this, we make a piecewise Chebyshev interpolant with two independent polynomials for $\ell \in [2, 10]$ and $\ell \in [10, 3000]$ (Limber approximation off/on).
This lowers the Chebyshev interpolation error to the level of cubic splines.

Chebyshev interpolation is also simpler.
Fixing the computational cost to exactly 70 $\ell$-points,
we first tried a cubic spline with a uniform $\ell$-grid,
but this led to large interpolation errors for small $\ell$.
After progressively refining the $\ell$-grid with trial and error,
we ended up with a high-medium-low density $\ell$-grid over the subintervals of $\ell \in [2, 10, 100, 3000]$.
Meanwhile, Chebyshev interpolation performs better with a straightforward $\ell$-grid.

The figure also shows that integer-rounded and standard Chebyshev $\ell$-interpolation perform equally well.

Here we concentrated on a flat cosmology with spherical Bessel functions.
In curved universes they generalize to hyperspherical Bessel functions \citep[e.g.][]{lesgourguesFastAccurateCMB2014a}.
They are harder to compute, typically using recurrence relations between integer $\ell$ \citep{tramComputationHypersphericalBessel2017a}
instead of solving the ODE \eqref{eq:besselode}, which receives an extra curvature-dependent term.
We therefore anticipate that the integer-rounded Chebyshev interpolation strategy would be easiest to generalize to curved geometries.

Thus, Chebyshev interpolation is better also in $\ell$.
It reaches up to several orders of magnitude lower error in CMB spectra than cubic splines with the same number of samples.
The emulator Capse.jl for $C_\ell(p)$ of cosmological parameters $p$ also uses it,
but their main purpose is to train on fewer independent coefficients $a_m(p)$ in the Chebyshev series $C_\ell(p) = \sum_m a_m(p) T_m(x(\ell))$ instead of every $C_\ell(p)$ \citep{boniciCapsejlEfficientAutodifferentiable2024a}.
\Cref{sec:coeffs} links the coefficients to our barycentric approach.

\section{Interpolation in wavenumber and multipole}
\label{sec:kl}

\begin{figure*}
	\centering
	\includegraphics[width=\textwidth]{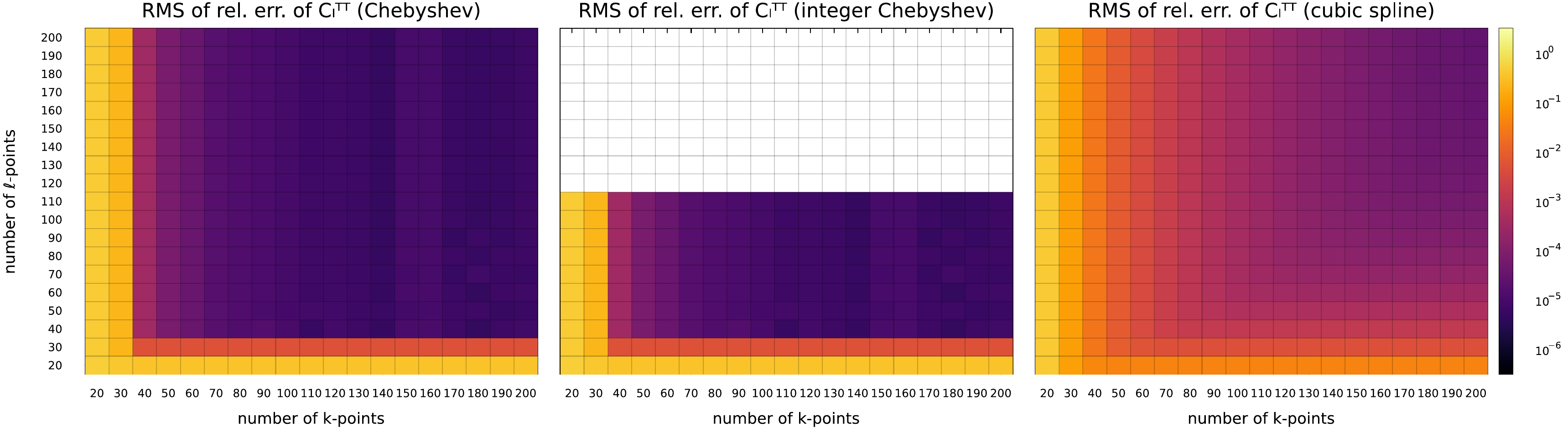}
	\caption{%
		\label{fig:kl}
		Root-mean-square of the relative error $\big(\sum_\ell (C_\ell^\mathrm{TT}/\bar{C}_\ell^\mathrm{TT}-1)^2/N_\ell\big)^\frac12$ of the CMB temperature power spectrum $C_\ell^\mathrm{TT}$
		for $50 \leq \ell \leq 2500$,
		computed with Chebyshev interpolation in both $k$ and $\ell$ versus equispaced cubic splines in both $k$ and $\ell$ with different numbers of samples.
		The error is relative to a reference spectrum $\bar{C}_\ell^\mathrm{TT}$ computed with Chebyshev interpolation from 1025 $k$-points and 513 $\ell$-points.
		The middle plot uses integer-rounded Chebyshev $\ell$-interpolation and is capped at 110 Chebyshev nodes,
		because a higher number would round several $\ell$-nodes to the same integer.
	}
\end{figure*}

In the previous section on $\ell$-interpolation,
we always used Chebyshev $k$-interpolation of the source functions $S(\tau, k)$ before line-of-sight integration in order to isolate the effect to $\ell$.

\Cref{fig:kl} instead shows the combined effect:
the CMB temperature spectrum $C_\ell^\mathrm{TT}$ interpolated with cubic splines in both $k$ and $\ell$,
versus Chebyshev interpolation in both $k$ and $\ell$.
Here we skip the cosmic variance-dominated $\ell \lesssim 50$ in \cref{fig:cmbTT},
and take the root-mean-square of the relative $C_\ell$ error for the smoothest $50 \leq \ell \leq 2500$ as a function of the number of $k$-points and $\ell$-points.
Chebyshev interpolation converges to the $10^{-4}$--$10^{-5}$-level with only 50--80 points in both $k$ and $\ell$,
while cubic splines approach the $10^{-4}$-level around 200 points.
This speeds up the perturbations and hence the full computation by $2.5\times$--$4\times$ for a fixed high precision level
(e.g. 0.2 s for $50 \times 50$ points vs 0.8 s for $200 \times 200$ on our machine).
Line-of-sight integration is sped up by a similar factor,
but is not the main bottleneck of the computation in flat universes.
In curved universes, hyperspherical Bessel functions are more expensive to compute,
so the speedup of line-of-sight integration may become more visible there.

\section{Conclusion}
\label{sec:conclusion}

We implemented interpolation with Chebyshev polynomials with the barycentric formula in the approximation-free \EB{} code SymBoltz.
By solving the perturbation ODEs and line-of-sight integrals at Chebyshev nodes in $k$ and $\ell$,
we reached significantly lower interpolation error than traditional cubic splines using the same number of samples.%
\footnote{Results in this paper were generated with SymBoltz 1.7.0 and the script \url{https://github.com/hersle/SymBoltz.jl/blob/v1.7.0/scripts/chebyshev.jl}.}

The rapid convergence of Chebyshev interpolation for smooth functions forms a synergy with approximation-free \EB{} codes.
In contrast, approximation-based codes switch equations in $\tau$, $k$ and $\ell$ to solve each point faster,
but complicate the physics and numerics and break smoothness.
In other words, Chebyshev interpolation provides approximation-free codes a unique way to win back performance and precision.

Standard Chebyshev interpolation in $\ell$ requires line-of-sight integrals and spherical Bessel functions generalized to non-integer $\ell$,
but we showed how to avoid this by rounding Chebyshev nodes to their nearest integer.
This near-Chebyshev strategy performs comparably and would be easier to use in curved universes,
which need hyperspherical Bessel functions that are typically computed with integer-based recurrences and polynomials.

If implemented carefully, we believe approximation-based codes could also benefit from Chebyshev interpolation.
As it is a global method,
imperfections at one point can slow interpolation convergence over the whole domain.
For example, we saw that the Limber approximation slowed convergence of Chebyshev $\ell$-interpolation,
but restored it by interpolating piecewise for $\ell < \ell_\text{Limber}$ and $\ell \geq \ell_\text{Limber}$.
The same strategy could be used for approximations in $k$-space,
but is not straightforward when the switching criterion varies with both $\tau$ and $k$.
Cubic splines are safer with approximations because their error decays faster around local non-smooth features,
but converge more slowly.

Interpolation is only one of many computational methods in \EB{} codes.
Our scope here was to study convergence only of interpolation,
while fixing other precision parameters.
Of course, lower precision of line-of-sight quadrature, the Limber approximation and other methods may overshadow the high precision of Chebyshev interpolation.
Convergence with respect to \emph{all} precision parameters calls for more research into high-precision methods,
such as better line-of-sight integration and beyond-Limber methods \citep[e.g.][]{schonebergTraditionalLineofSightApproach2018,fangLimberEfficientComputation2020,chiarenzaBLASTLimberAngular2024a,reischkePylevinEfficientNumerical2025}.

We believe Chebyshev interpolation is a step forward for simpler \EB{} codes
with numerical algorithms that make better use of the physical properties of the equations,
instead of resorting to complicated approximation schemes.

\begin{acknowledgements}
	I thank Hans A. Winther for providing valuable feedback on drafts of this paper.
	I thank Michael Helton for creating the Bessels.jl package,
	which made it easy to compute spherical Bessel functions with both integer and non-integer order in this paper.
	I thank Steven G. Johnson for creating the FastChebInterp.jl package,
	which made it easy to inspect the convergence of Chebyshev coefficients in this paper.
	This research was supported by the Research Council of Norway under project number 325113.
\end{acknowledgements}

\bibliographystyle{aa}
\bibliography{paper}

\appendix
\crefalias{section}{appendix} % make cleverref render "appendix A" instead of "section A"
\nolinenumbers

\section{Chebyshev coefficients}
\label{sec:coeffs}

Instead of interpolating with the barycentric formula in \cref{sec:chebyshev},
a function can be approximated with a series of Chebyshev polynomials.
This alternative approach to Chebyshev interpolation is particularly useful for error analysis.
We give a brief summary of this approach based on \cite{trefethenApproximationTheoryApproximation2019}.

The Chebyshev polynomials of the first kind are defined as
\begin{equation}
	T_n(x) = \cos(n \arccos x)
	\qquad
	(x \in [-1, 1]).
\label{eq:chebpolydef}
\end{equation}
To see it is really a polynomial,
evaluate $T_0(x)$ and $T_1(x)$ explicitly and use the identities $\cos(n\theta\pm\theta) = \cos(n\theta) \cos(\theta) \mp \sin(n\theta) \sin(\theta)$ with $\theta = \arccos x$ to get the polynomial recurrence%
\begin{equation}
	T_0(x) = 1,
	\quad
	T_1(x) = x,
	\quad
	T_{n+1}(x) = 2 x T_n(x) - T_{n-1}(x).
\label{eq:recurrence}
\end{equation}

The Chebyshev polynomials are an orthogonal basis for expanding a function $f(x)$ on $x \in [-1, 1]$ in a unique series%
\begin{equation}
	f(x) = \sum_{m=0}^\infty a_m T_m(x).
\label{eq:infseries}
\end{equation}
They are orthogonal with respect to the inner product
(change variables to $x = \cos \theta$ and use orthogonality of cosines)
\begin{equation}
	\int_{-1}^1 \frac{T_m(x) T_{m'}(x)}{\sqrt{1-x^2}} \, \diff{x} =
	\begin{cases} 0 & \text{if $m \neq m'$,} \\ \pi & \text{if $m=m'=0$,} \\ \pi/2 & \text{if $m=m' \geq 1$,} \end{cases}
\label{eq:innerprodcont}
\end{equation}
so the coefficients that represent a given function $f(x)$ are
\begin{equation}
	a_m = \begin{cases}
		\displaystyle \frac{1}{\pi} \int_{-1}^1 \frac{f(x) T_0(x)}{\sqrt{1-x^2}} \, \diff{x} & \text{if $m = 0$}, \\
		\displaystyle \frac{2}{\pi} \int_{-1}^1 \frac{f(x) T_m(x)}{\sqrt{1-x^2}} \, \diff{x} & \text{if $m \geq 1$}.
	\end{cases}
\label{eq:coeffscont}%
\end{equation}
These coefficients represent $f(x)$ in the infinite-dimensional space \eqref{eq:infseries},
and are not very practical.

\subsection{Chebyshev projection}

One way to approximate $f(x)$ is to take the coefficients \eqref{eq:coeffscont},
but truncate the infinite series \eqref{eq:infseries} at a finite $n$:
\begin{equation}
	f(x) \approx \hat{p}_n(x) = \sum_{m=0}^n a_m T_m(x).
\label{eq:projection}
\end{equation}
This is called Chebyshev projection,
because dropping higher-order coefficients effectively projects $f(x)$ from an infinite-dimensional space of continuous functions
onto a finite lower-dimensional space of polynomials.

Importantly, this is not interpolation because it does not pass through a given sample of function values.
Instead it deals with the integrals \eqref{eq:coeffscont}, which need $f(x)$ to be evaluated at arbitrary $x$.
This can be appropriate when one has an expression for $f(x)$ and can evaluate the integrals analytically,
but it is impractical for other purposes.

\subsection{Chebyshev interpolation}

% https://www.buttenschoen.ca/MATH551/build/point-choice-4c36fbc36ea8e8d082339f77a23601f8.pdf
A practical interpolation method is to construct the polynomial $p_n(x)$ that passes through $f(x)$ at certain points $x_i$.
However, instead of expanding it in Lagrange polynomials as in \cref{sec:chebyshev},
we can expand it in a series of Chebyshev polynomials:
\begin{equation}
	f(x) \approx p_n(x) = \sum_{m=0}^n c_m T_m(x)
	\quad \text{such that} \quad
	p_n(x_i) = f(x_i).
\label{eq:interpolation}
\end{equation}

It is convenient to select $x_i$ as the extrema of $T_n(x)$.
The trigonometric definition \eqref{eq:chebpolydef} reaches $\lvert T_n(x) \rvert = 1$ when $n\theta = i\pi$ for integer $i$,
so its extrema are
\begin{equation}
	x_i = \cos \left( \frac{i}{n} \pi \right)
	\qquad
	(0 \leq i \leq n).
\label{eq:chebnodes2_app}
\end{equation}
Analogously to the continuous inner product \eqref{eq:innerprodcont},
the extrema give rise to a discrete orthogonality relationship:
\begin{equation}
	\sideset{}{''}\sum_{i=0}^n T_m(x_i) T_{m'}(x_i) =
	\begin{cases} 0 & \text{if $m \neq m'$,} \\ n & \text{if $m = m' \in \{0, n\}$,} \\ n/2 & \text{if $m = m' \in [1,n-1]$,} \end{cases}
\label{eq:innerproddiscrete}
\end{equation}
where the double-primed sum halves its first and last terms.
Thus, the coefficients $c_m$ that interpolate a given $f(x)$ are%
\begin{equation}
	c_m = \begin{cases}
		\displaystyle \frac{1}{n} \sideset{}{''}\sum_{i=0}^n f(x_i) \cos \left( \frac{im\pi}{n} \right) & \text{if $m \in \{0, n\}$}, \\
		\displaystyle \frac{2}{n} \sideset{}{''}\sum_{i=0}^n f(x_i) \cos \left( \frac{im\pi}{n} \right) & \text{if $m \in [1, n-1]$}.
	\end{cases}
\label{eq:coeffsdiscrete}
\end{equation}
This is a type-I Discrete Cosine Transform (DCT).
It can be computed with only $O(n \log n)$ operations.
This special property makes Chebyshev interpolation particularly attractive.

Once $c_m$ are known, the interpolating series \eqref{eq:interpolation} is easily evaluated for any $x$ with $O(n)$ operations.
Forward iteration through the recurrence \eqref{eq:recurrence} may be most natural, but this is unstable for $\lvert x \rvert \rightarrow 1$.
A stable way is to iterate backwards with the \cite{clenshawNoteSummationChebyshev1955} algorithm with an intermediate variable $b_m(x)$:%
\begin{subequations}
\begin{align}
	b_{n+1}(x) &= b_{n+2}(x) = 0, \\
	b_m(x) &= c_m + 2x b_{m+1}(x) - b_{m+2}(x) \quad (m = n,\, \ldots,\, 0), \\
	p_n(x) &= \textstyle{\frac12 \big(c_0 + b_0(x) - b_2(x)\big)}.
\end{align}
\end{subequations}

Interpolating with the series \eqref{eq:interpolation} is equivalent to the barycentric formula \eqref{eq:bary}
with the Chebyshev points \eqref{eq:chebnodes2} and weights \eqref{eq:chebweights2} of the second kind,
as the polynomial is of the same order and goes through the same points.
The DCT \eqref{eq:coeffsdiscrete} converts the function values $f(x_i)$ of the barycentric approach to the coefficients $c_m$ of the series approach.
While the barycentric approach is simpler, faster and more flexible,
the advantage of the series is that the coefficients
provide a useful diagnostic for the maximum interpolation error \citep[Chapter 4]{trefethenApproximationTheoryApproximation2019}:
\begin{equation}
	\big\lvert f(x) - p_n(x) \big\rvert \leq 2 \sum_{m=n+1}^\infty \lvert a_m \rvert .
\label{eq:coefferr}
\end{equation}
This bound involves the projection coefficients $a_m$, not the interpolation coefficients $c_m$.
The two differ by aliasing effects that diminish as the series converges.
Thus $c_m \rightarrow a_m$ in practice,
so we can compute the DCT coefficients \eqref{eq:coeffsdiscrete} and check for their convergence,
which also implies convergence of the full $p_n(x)$.

SymBoltz always interpolates with the barycentric formula.
In addition,
we added a way to compute the DCT for the coefficients \eqref{eq:coeffsdiscrete} with the FastChebInterp.jl\footnote{\url{https://github.com/JuliaMath/FastChebInterp.jl}} package,
as an additional way to inspect the interpolation convergence in \cref{fig:coeffs}.

\end{document}